\documentclass[twocolumn]{aastex631}
\usepackage{graphicx}
\usepackage{tablefootnote}

\received{\today}

\submitjournal{ApJ}

\begin{document}

\title{The magnetic field environment of active region 12673 that produced the energetic particle events of September 2017}

\correspondingauthor{Stephanie L. Yardley}
\email{stephanie.yardley@ucl.ac.uk}

\author[0000-0003-2802-4381]{Stephanie L. Yardley}
\affiliation{University College London, Mullard Space Science Laboratory, Holmbury St. Mary, Dorking, Surrey, RH5 6NT, UK}
\affiliation{Donostia International Physics Center (DIPC), Paseo Manuel de Lardizabal 4, 20018 San Sebasti{\'a}n, Spain}

\author[0000-0002-0053-4876]{Lucie M. Green}
\affiliation{University College London, Mullard Space Science Laboratory, Holmbury St. Mary, Dorking, Surrey, RH5 6NT, UK}

\author[0000-0001-7927-9291]{Alexander W. James}
\affiliation{European Space Agency (ESA), European Space Astronomy Centre (ESAC), Camino Bajo del Castillo s/n, 28692 Villanueva De La Ca{\~n}ada, Madrid, Spain}

\author[0000-0002-1365-1908]{David Stansby}
\affiliation{University College London, Mullard Space Science Laboratory, Holmbury St. Mary, Dorking, Surrey, RH5 6NT, UK}
\affiliation{University College London/Research IT Services, Gower St, Bloomsbury, London WC1E 6BT, UK}

\author[0000-0001-8055-0472]{Teodora Mihailescu}
\affiliation{University College London, Mullard Space Science Laboratory, Holmbury St. Mary, Dorking, Surrey, RH5 6NT, UK}

\begin{abstract}
Forecasting solar energetic particles (SEPs), and identifying flare/CMEs from active regions (ARs) that will produce SEP events in advance is extremely challenging. We investigate the magnetic field environment of AR 12673, including the AR's magnetic configuration, the surrounding field configuration in the vicinity of the AR, the decay index profile, and the footpoints of Earth-connected magnetic field, around the time of four eruptive events. Two of the eruptive events are SEP-productive (2017 September 4 at 20:00~UT and September 6 at 11:56~UT), while two are not (September 4 at 18:05~UT and September 7 at 14:33~UT). We analysed a range of EUV and white-light coronagraph observations along with potential field extrapolations and find that the CMEs associated with the SEP-productive events either trigger null point reconnection that redirects flare-accelerated particles from the flare site to the Earth-connected field and/or have a significant expansion (and shock formation) into the open Earth-connected field. The rate of change of the decay index with height indicates that the region could produce a fast CME ($v >$ 1500~km~s$^{-1}$), which it did during events two and three. The AR's magnetic field environment, including locations of open magnetic field and null points along with the magnetic field connectivity and propagation direction of the CMEs play an important role in the escape and arrival of SEPs at Earth. Other SEP-productive ARs should be investigated to determine whether their magnetic field environment and CME propagation direction are significant in the escape and arrival of SEPs at Earth.

\end{abstract}


\shorttitle{Magnetic environment of SEP-productive AR12673}
\shortauthors{Yardley et al.}


\section{Introduction}  \label{sec:intro}

The Sun sporadically accelerates particles (electrons, protons and heavy ions) to near-relativistic speeds and energies of 10 keV to GeV during activity events such as solar flares and coronal mass ejections (CMEs). These particles are termed solar energetic particles (SEPs). The acceleration processes are thought to be related to the electric fields or plasma turbulence associated with the magnetic reconnection involved in solar flares that energise the thermal plasma to suprathermal levels \citep[see][for a review]{vlahos2019}. For example, SEP production is correlated with flare thermal energy, with all flares with a GOES soft X-ray classification greater than X5, and located in the solar longitude range W15 to W75, being SEP productive \citep{belov2005}.

The most energetic flares have a high likelihood of being accompanied by a CME \citep{yashiro2005} and correlations have been found between certain CME characteristics and SEP production. For example, CME acceleration and spatial extent have been shown to influence the production and spread of SEPs \citep{kahler1986} and CME energy is correlated to peak SEP intensity \citep{kahler2013}. The mechanism by which CMEs are capable of producing SEPs is through the shocks that are created by super-Alfv{\'e}nic CMEs as they move through the lower corona and into the solar wind \citep{reames1999,Kahler2001, gopalswamy2004, papaioannou2016}, injecting shock-accelerated particles onto observer-connected field lines. 

Regardless of the SEP acceleration mechanism, accelerated particles escape the Sun by propagating along open field lines that guide the particles as they move through the heliosphere. SEP studies utilising data from spacecraft at different solar longitudes have shown that the largest SEP intensities are detected at those spacecraft that are well connected to the solar activity event \citep{dresing2014, lario2013}. Therefore, Earth impacting SEPs must either originate on, diffuse onto, or have access to as a result of magnetic reconnection, magnetic field lines that connect from the Sun to the Earth. Currently, the source of SEP seed populations, the method by which particles escape from their acceleration region, and the SEP profile variation from event to event in relation to source region characteristics are still not yet well understood \citep{desai2016}. For example, the standard CSHKP model of solar flares \citep{carmichael1964, sturrock1966, hirayama1974, kopp1976} injects reconnection-accelerated particles downward to produce heating in the lower atmospheric plasma and the flare, but also injects particles upward. However, these upward-accelerated particles are injected onto the closed field lines of the escaping CME and therefore the escape routes for the SEPs are not easily explained.

It is possible to investigate the escape of SEPs by first identifying their solar origin. A key technique  for this involves analysing the elemental composition of the SEP plasma population as measured in situ and then seeking plasma with the same composition in the flare/CME source region. Elemental composition is normally characterised by considering elements with differing first ionisation potentials (FIP), comparing low-FIP to high-FIP elements, such as Si and S, and comparing coronal abundances to that of the photosphere. This has emerged as a key diagnostic in the study of SEP plasma \citep{reames2018}. Recently, this approach has been used to trace multiple SEP events detected near Earth during January 2014 back to their solar source \citep{brooks2021a}. The results showed that plasma confined by strong magnetic fields in the active region (AR) developed the composition signature (high Si/S abundance ratio) indicative of the SEP population. Smaller Si/S abundance enhancements were also recorded close to upflow regions at the AR boundary. The plasma detected in situ during the SEP events was therefore determined to be a combination of plasma that was accelerated and released during the flare/CME itself, that escaped directly along open magnetic field lines, and also plasma that escaped indirectly through interchange reconnection at the AR periphery \citep{brooks2021b,yardley2021}.

It is clear that the configuration of an AR's magnetic field, and the configuration of its surroundings, plays a key role in both particle acceleration and escape. The magnetic field configuration influences where flare reconnection may occur, how much energy can be released, and over what timescale. The magnetic field configuration also affects CME acceleration, speed and propagation direction. Therefore, the magnetic configuration of an AR and its surrounding field must be investigated in SEP studies. Determining whether there are certain characteristics of an AR's magnetic field or of the surrounding field that are necessary for SEP-production and escape would therefore mark a step towards being able to forecast which regions might produce observer-impacting SEP events.

For the specific case of CMEs that are initiated by an unstable flux rope \citep{kliem2006}, the pre-eruptive (and therefore stable) configuration is obtained when the upward Lorentz force of the rope is balanced by the downward strapping force of the overlying arcade. The flux rope will become unstable if it reaches a height at which the downward force of the overlying strapping is insufficient. This height is known as the critical height and the gradient of the strapping field is given by the decay index \citep{kliem2006}. Once unstable, the decay index profile will influence the acceleration of the CME, and its terminal velocity. A study by \citet{kliem2021} suggests there is a correlation between the steepness of the decay index height profile (above the critical height) and the CME velocity for CMEs with speeds $\geqslant$1500~km~s$^{-1}$.

In this paper, we focus on the evolution of the well-studied NOAA AR 12673 during its disk passage \citep{sun2017, yang2017, chertok2018b, cohen2018, luhmann2018, romano2018, sharykin2018, shen2018, wang2018, anfinogentov2019, bruno2019, romano2019}, which was the site of several M and X-class flares, and fast CMEs between 2017 September 4 until it passed out of view over the west limb on 2017 September 10. Two SEP events and a further SEP event that produced a ground-level enhancement were produced in association with this activity, as described in \cite{bruno2019}. Two SEP events occurred while the AR was visible on disk from the Earth perspective, and within 50W of central meridian. We aim to determine the role of the magnetic field environment of the AR and its surroundings in enabling these two flare/CME events to produce SEP events that were detected at Earth, whereas other major flares/CMEs from the AR were not. We probe whether there are certain characteristics of the AR magnetic field configuration, and its surrounding magnetic environment, that influence the production and escape of SEPs in a subset of the major flare/CME events produced by the AR. In contrast to previous studies, we analyse both the SEP and non-SEP productive events.


\section{Evolution of NOAA active region 12673}

\begin{figure*}
\centering
\includegraphics[width=1.0\textwidth]{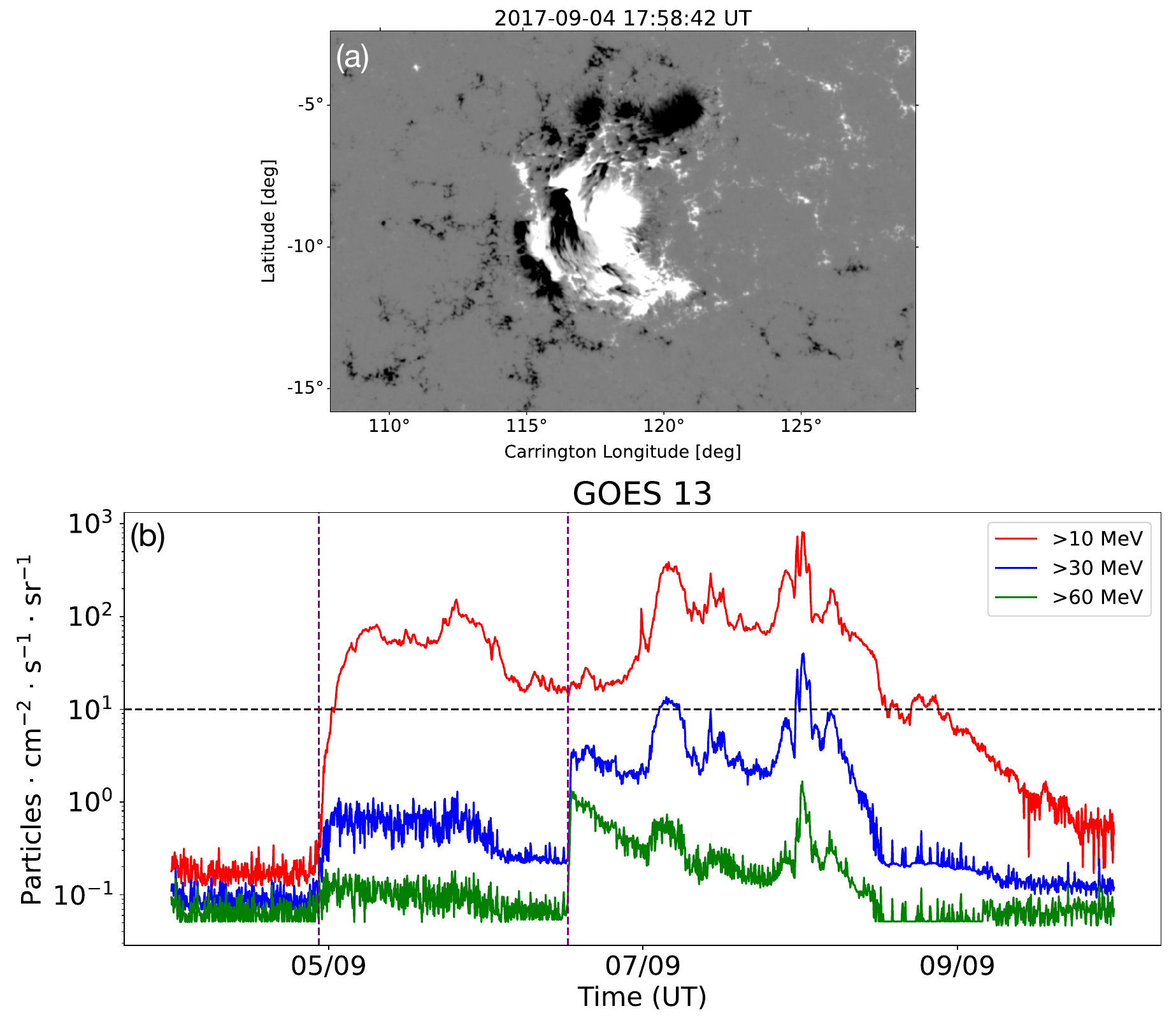}
\caption{Panel (a) shows the radial photospheric magnetic field of AR 12673  taken by the Helioseismic Magnetic Imager on board the Solar Dynamics Observatory (SDO/HMI; \citealt{scherrer2012}). The radial magnetic field component is from the HMI SHARP data series (Spaceweather HMI Active Region Patch; \citealt{bobra2014}) where positive (negative) field is represented by white (black), respectively, saturated at $\pm$500~G. Panel (b) displays the GOES-13 proton flux (10, 30, 60 MeV channels) with two SEP events detected on 2017 September 4 and 6. The black dashed line represents the 10 MeV warning threshold above which a proton event alert is issued by NOAA/SWPC. The purple dashed lines indicate the start times of the two SEP events, which are 2017 September 4 at 22:30~UT and 2017 September 6 at 12:35~UT, respectively.}
\label{hmi_goes}
\end{figure*}

NOAA active region (AR) 12673 appeared on the east solar limb on 28 August 2017, consisting of a lone positive polarity sunspot with dispersed positive (negative) polarity field to the north-west (south-east). This positive polarity spot was present in previous rotations as part of AR 12670 and AR 12665 (which also produced two SEP events). Major flux emergence began in the region on 2017 September 2 and as a consequence, the AR evolved rapidly, from an $\alpha$ to a $\beta\gamma\delta$ Hale-class by September 5 (Figure~\ref{hmi_goes}~a). The region was recorded to have had one of the fastest rates of emerging flux ever observed \citep{sun2017}. 

The AR first started flaring early on 2017 September 4 and produced its first CME later that day, which was observed to begin around 18:00~UT. In total, AR 12673 was the source of 27 GOES M- and four X-class flares, eleven CMEs and three SEP events during the time period 2017 September 4--10. Figure~\ref{hmi_goes}~(b) shows the two SEP events that occurred as detected by GOES when the AR was less than 50W of central meridian.  

In the next section, we focus on analysing the properties of a subset of eruptive flares from this AR (between 2017 September 4 and 7), which are both SEP-productive and non-SEP productive, in order to try to distinguish these two types of eruptive events.

\section{SEP and non-SEP flare/CME productive events} \label{sec:events}

In this study, we focus on four eruptive events (flares and their associated CMEs) that occur between 2017 September 4 and 7, when the AR was no greater than 50W of central meridian. Details of the flare and CME properties are given in Table~\ref{tab:summary_table}. The SEP productive events (events two and three) are temporally associated with an M5.5 GOES class flare on 2017 September 4 and a X9.3 GOES class flare on 2017 September 6. GOES 13 and 15 data show that particles from September 4 SEP event arrived at the spacecraft by 22:30~UT on the same day and on September 6 the particles of the SEP event were detected by 12:35 UT. Energy spectra of both SEP events were relatively soft, with the data from the 2017 September 4 event suggestive of a post-eruption origin \citep{chertok2018a,chertok2018b}. In the following, a brief description of each eruptive event is given, including the location and development of flare ribbons (``active PILs"), CME propagation direction and radial speed, and sites of magnetic reconnection as evidenced by the flaring and EUV observations. In total, four active PILs are identified in the AR along which flare ribbons are observed, three of which are aligned almost north-south (PILs 1, 2, and 3 in Figure~\ref{Event1}~b) and one aligned east-west (PIL 4). The AR evolves so that on September 7 only one active PIL is remaining (PIL 3).

\subsection{Eruptive Event 1}
Event one occurred on 2017 September 4 and comprises of an M1.0 GOES class flare that occurred in association with a CME. No SEPs were detected in association with this event. Flare ribbons are initially observed along PIL 2, but an overlap with ribbons along this PIL from a flare just minutes before hinders the ability to discriminate which flare is responsible for these ribbons. As seen in AIA 1600~\AA, flare ribbons are forming along PILs 3 and 4 by 18:16 UT on 2017 September 4, which is temporally coincident with the initial increase in the GOES soft X-ray light curve for this flare (see Figure~\ref{Event1} panels a and b). At the peak of the flare, as determined from the GOES soft X-ray emission (18:21~UT), the flare ribbons are mainly observed along PIL 4. The eruption begins at around $\sim$18:05~UT as observed by the expansion and propagation of coronal loops to the south-west, visible in the SDO/AIA 171~\AA~and running difference images (Figure~\ref{Event1}~c). The CME (Figure~\ref{Event1}~d) is first seen in LASCO/C2 data at 19:00~UT on September 4 and has a radial speed of 973~km~s$^{-1}$ (calculated using the STEREOCAT tool\footnote{\url{https://ccmc.gsfc.nasa.gov/analysis/stereo/}}). The CME propagation direction in 3D is S06W28 (taken from the DONKI catalogue\footnote{\url{https://kauai.ccmc.gsfc.nasa.gov/DONKI/}}), which is to the west of the radial direction of the AR.

\subsection{Eruptive Event 2}  
Event two also occurs on 2017 September 4 and comprises of an M5.5 GOES class flare that occurred in association with a CME and the production of SEPs. Flare ribbons are first observed (faintly) at $\sim$20:11 UT on 2017 September 4 in the AIA 1600~\AA~waveband, initially appearing along PIL 1. The ribbons then spread across PILs 4, 3 and then 2. At the peak of the flare's soft X-ray emission (20:32~UT) the 1600~\AA~waveband flare ribbons are most intense across PIL 2 (see Figure~\ref{Event2} panels a and b). Reverse S-shaped coronal loops are observed to erupt to the north-west from 20:00~UT onwards, which drives reconnection at what looks like a null point, as evidenced by the 171~\AA~and running difference images, located north-east of the AR (Figure~\ref{Event2}~c). The associated CME is first seen in LASCO/C2 (appearing to the south) on 2017 September 4 at 20:36~UT. The eruption direction in 3D is S10W10 (from DONKI), which is north of the radial direction of the AR, and is super-imposed on the previous CME from the region that occurred during event one (Figure~\ref{Event2}~d). However, the CME is also observed to have a component propagating to the east as seen in the coronagraph field-of-view. The CME's radial speed is calculated to be 2153~km~s$^{-1}$ (from the STEREOCAT tool). A type II radio burst was observed in association with this event (as recorded by the WIND/WAVES catalogue\footnote{\url{https://cdaw.gsfc.nasa.gov/CME\_list/radio/waves\_type2.html}} \citealt{gopalswamy2019}). The SEP event is first detected in the GOES data at 22:30~UT.

\startlongtable
\begin{deluxetable*}{cccccccccccc}
\tabletypesize{\scriptsize}
\tablecolumns{12}
\tablewidth{0pt}
\tablecaption{The flare/CME properties of the four eruptive events. \label{tab:summary_table}}
\tablehead{
\colhead{No.}  & \colhead{Lat.} & \colhead{Lon.} & \colhead{Flare} & \colhead{Flare} & \colhead{GOES} & \colhead{CME} & \colhead{LASCO/C2} & \colhead{Half} & \colhead{Radial} & \colhead{Prop.}  & \colhead{SEP} \\ & \colhead{(deg)} & \colhead{(deg)} & \colhead{Start Time} & \colhead{Peak Time} & \colhead{Flare} & \colhead{Onset Time} & \colhead{First Obs.} & \colhead{Width} & \colhead{Velocity} & \colhead{Dir.} &  \colhead{Event} \\  & & & \colhead{(UT)} & \colhead{(UT)} & \colhead{Class} & \colhead{(UT)} & \colhead{(UT)} & \colhead{(deg)} & \colhead{(km~s$^{-1}$)} & & }
\startdata
1 & -7 & 11 & 17 Sep 4 18:12  & 17 Sep 4 18:21 & M1.0 & 17 Sep 4 18:05 & 17 Sep 4 19:00 & 37 & 973 & S06W28 & N \\
2 & -10 & 11 & 17 Sep 4 20:12 & 17 Sep 4 20:32 & M5.5 & 17 Sep 4 20:00 & 17 Sep 4 20:36 & 101 & 2153 & S10W10 & Y \\
3 & -9 & 34 & 17 Sep 6 11:52 & 17 Sep 6 12:01 & X9.3 & 17 Sep 6 11:56 & 17 Sep 6 12:24 & 103 & 2268 & S15W23 & Y \\
4 & -8 & 48 & 17 Sep 7 14:31 & 17 Sep 7 14:36 & X1.3 & 17 Sep 7 14:33 & 17 Sep 7 15:12 & 16 & 481 & S16W53 & N \\
\enddata
\tablecomments{The first three columns give the event number (1--4), the latitude and longitude that the eruptive event originates from. Columns four and five give the start and peak time of the solar flares as derived from the GOES soft X-ray flux. Columns six and seven give the time of the CME onset as observed in the SDO/AIA 171~\AA~data, and the time the CME was first observed in SoHO LASCO/C2. Columns eight and nine give the half width and the radial velocity as determined by the STEREOCAT tool. The ensemble mode is used and the median of 5 different speed measurements, calculated using LASCO/C2 and STEREO-A/COR2, is taken. The propagation direction of the CME is given in column ten, which is taken from the DONKI catalogue. Finally, column eleven states whether the event was associated with SEPs as detected by GOES.}
\end{deluxetable*}

\subsection{Eruptive Event 3}  
Event three occurred on 2017 September 6 and comprises an X9.3 GOES class flare that occurred in association with a CME and an SEP event. Flare ribbons are first observed $\sim$11:53 UT on September 6 in AIA 1600~\AA~waveband data along PIL 3 and then PIL 4. By September 6 the AR has evolved to have two main PILs (3 \& 4, see Figure~\ref{Event3}~b). Flare ribbons appear and are strongest around PIL 3 at the peak of the flare (12:01~UT) but also spread to PIL 4 (Figure~\ref{Event3}~a). The eruption associated with the flare, as observed in EUV imaging data of the lower corona, is quite complex. Three different structures are observed to propagate outwards from the AR (black arrows shown in Figure~\ref{Event3}~c). The first is a loop structure, aligned north south with respect to the AR, located at the east of the AR. The second and third loop structures originate from the western side of the AR. These loop structures (2 \& 3), which erupt around the same time as the first loop structure, propagate to the south-west and north-west, respectively (see Figure~\ref{Event3}~c). As a result, a halo CME was observed, with a radial speed of 2268~km~s$^{-1}$, which was first observed in LASCO/C2 at 12:24~UT on September 6. The CME propagated in the direction S15W23, which is to the east of the radial direction from the AR. A type II burst was observed in association with this event. The SEP event is first detected in the GOES data at 12:35~UT.
    
\subsection{Eruptive Event 4}
Event four begins on September 7 at 14:31~UT and is not associated with an SEP event. The AR is almost at 50W at this time with projection effects starting to become more evident. By this time, the AR has evolved to have only one main PIL (3, Figure~\ref{Event4}~b), which has changed orientation mainly due to shearing motions. The flare ribbons are therefore observed along PIL 3 and do not evolve into multiple ribbons (Figure~\ref{Event4}~a). The erupting loop structure is difficult to identify directly in this case but there are nearby loops that visibly oscillate due to the propagation of an erupting structure (Figure~\ref{Event4}~c). The eruption is indirectly observed to begin around 14:33~UT. A narrow CME is first observed in LASCO/C2 at 15:24~UT with a radial speed of 481 km~s$^{-1}$. The CME propagation direction is S16W53, which is to the south-west of the AR radial.

\begin{figure*}
\centering
\includegraphics[width=1.0\textwidth]{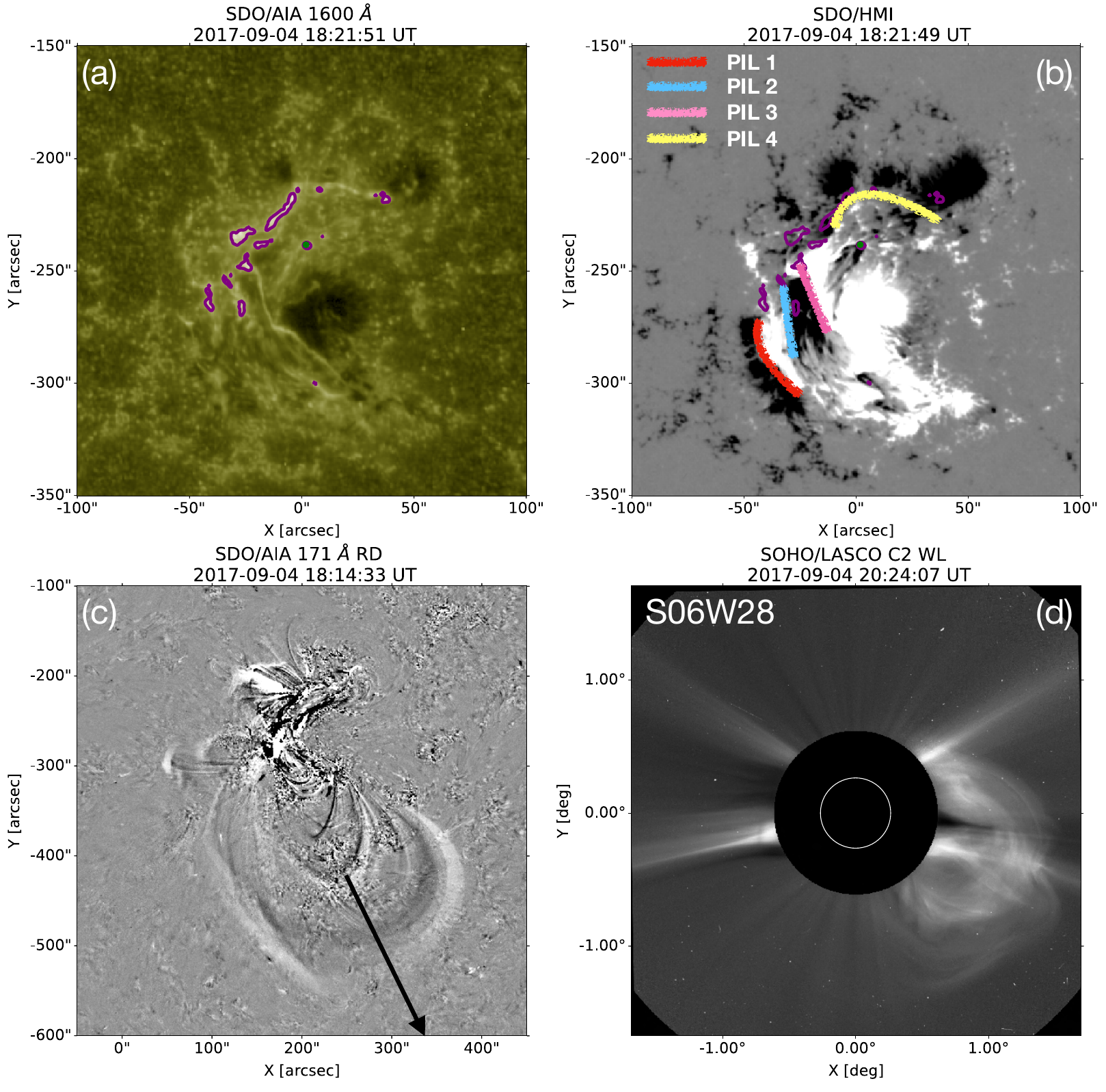}
\caption{Event one. Panel (a) shows the 1600~\AA~data taken by the Atmospheric Imaging Assembly on board SDO (SDO/AIA; \citealt{lemen2012}) at the peak time of the flare. The purple contours show the regions where the 1600~\AA\~emission of the flare ribbons is above a threshold of 20~\% of the maximum value. The green dot shows the location of the peak emission within the flare ribbon contours. Panel (b) shows the corresponding magnetogram of the longitudinal magnetic field (hmi.M\_45s data series) where white (black) represents positive (negative) field saturated at $\pm$500~G. The four PILs are indicated with the red, blue, pink and yellow lines representing PILs 1, 2, 3, and 4, respectively. Panel (c) shows a running difference SDO/AIA 171~\AA~image to reveal the erupting loop structures during the eruption. The online animation of panel (c) shows a movie of the SDO/AIA 171~\AA~running difference images with a 2~minute cadence between 17:00 and 18:58~UT on 2017 September 4 during the time period in which eruptive event one occurs. The black arrow shows the direction of travel of the erupting loop structures. Panel (d) displays a white light image of the CME taken by the Large Angle and Spectrometric COronagraph (LASCO/C2; \citealt{brueckner1995}) on board the Solar and Heliospheric Observatory (SoHO; \citealt{domingo1995}), where the CME propagation direction (from the DONKI catalogue) is given in the top left.
}
\label{Event1}
\end{figure*}

\begin{figure*}
\centering
\includegraphics[width=1.0\textwidth]{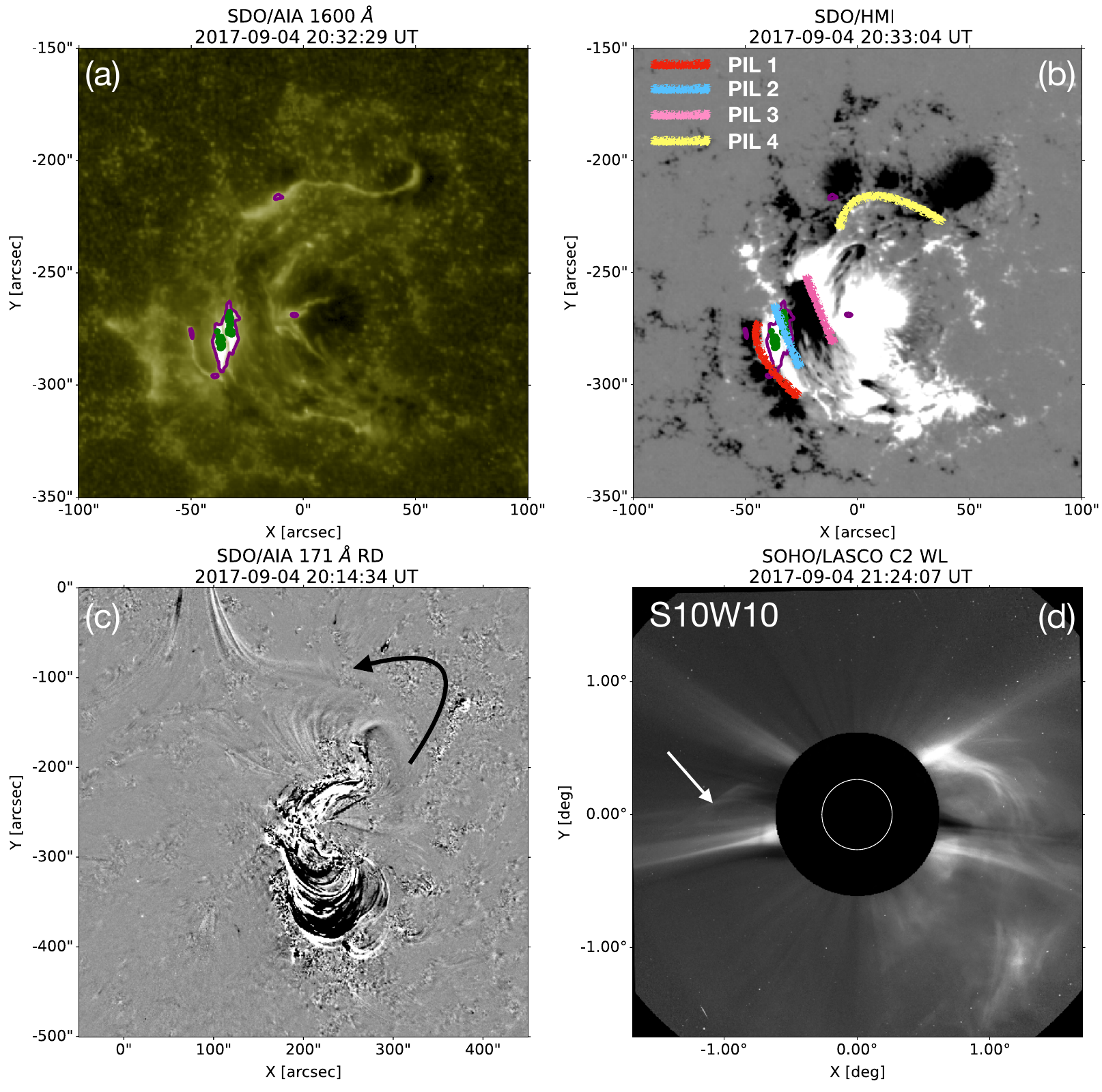}
\caption{The same as Figure~\ref{Event1} but for event two. The black arrow in panel (c) demonstrates the propagation direction of an erupting loop structure into what appears to be a null point. The online animation of panel (c) shows a movie of the SDO/AIA 171~\AA~running difference images with a 2~minute cadence between 19:33 and 21:29~UT on 2017 September 4 during the time period in which eruptive event two occurs. The white arrow in panel (d) indicates the eastern component of the erupting CME.
}
\label{Event2}
\end{figure*}

\begin{figure*}
\centering
\includegraphics[width=1.0\textwidth]{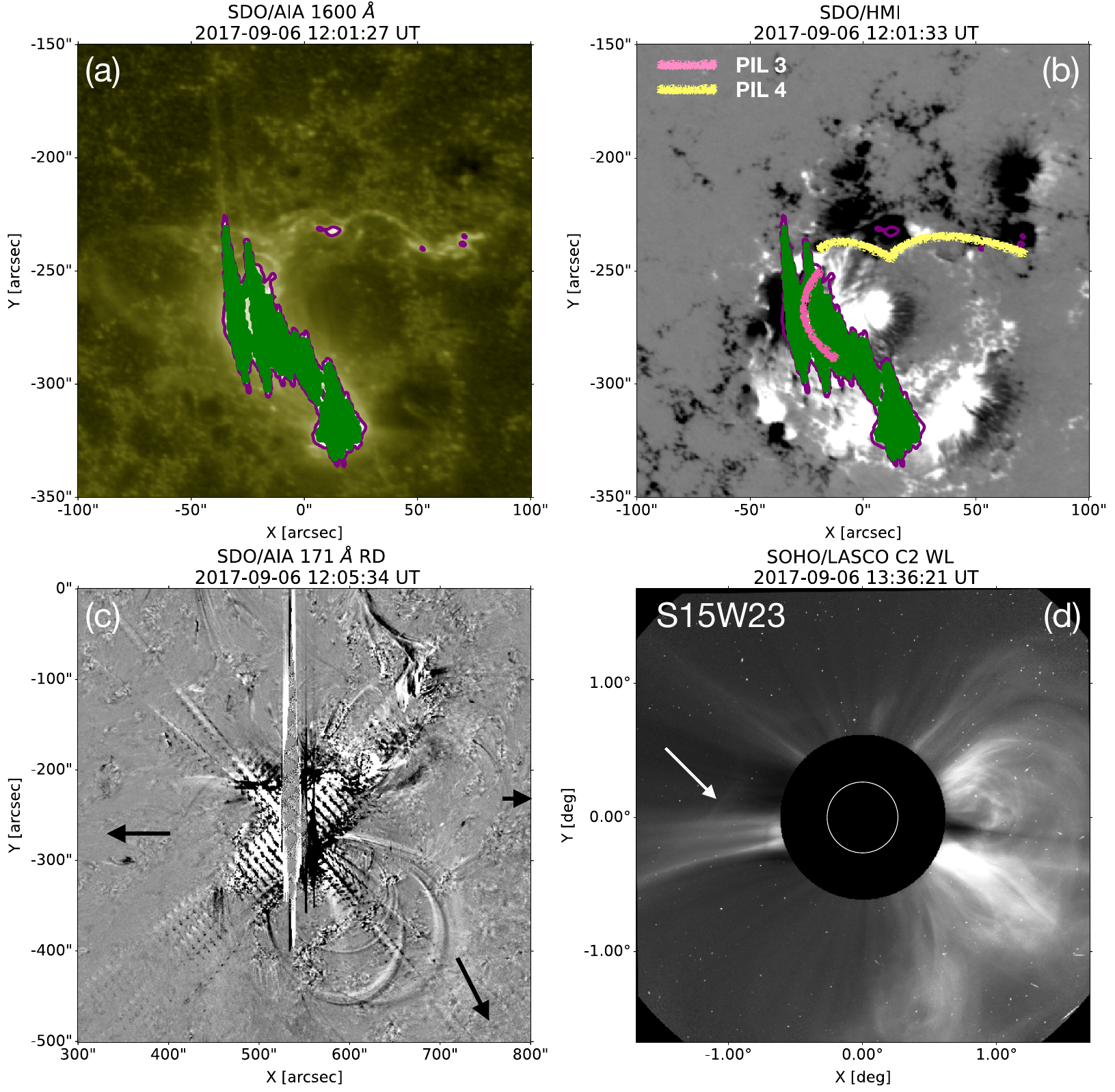}
\caption{Same as Figure~\ref{Event1} but for event three. Panel (b) shows that the AR has evolved so only two of the PILs (3 \& 4) are now present. In panel (c) the three black arrows indicate the three different erupting structures observed in the difference imaging. The online animation of panel (c) shows a movie of the SDO/AIA 171~\AA~running difference images with a 2~minute cadence between 11:31 and 13:29~UT on 2017 September 6 during the time period in which eruptive event three occurs. The white arrow in panel (d) shows the eastern component of the propagating CME. 
}
\label{Event3}
\end{figure*}

\begin{figure*}
\centering
\includegraphics[width=1.0\textwidth]{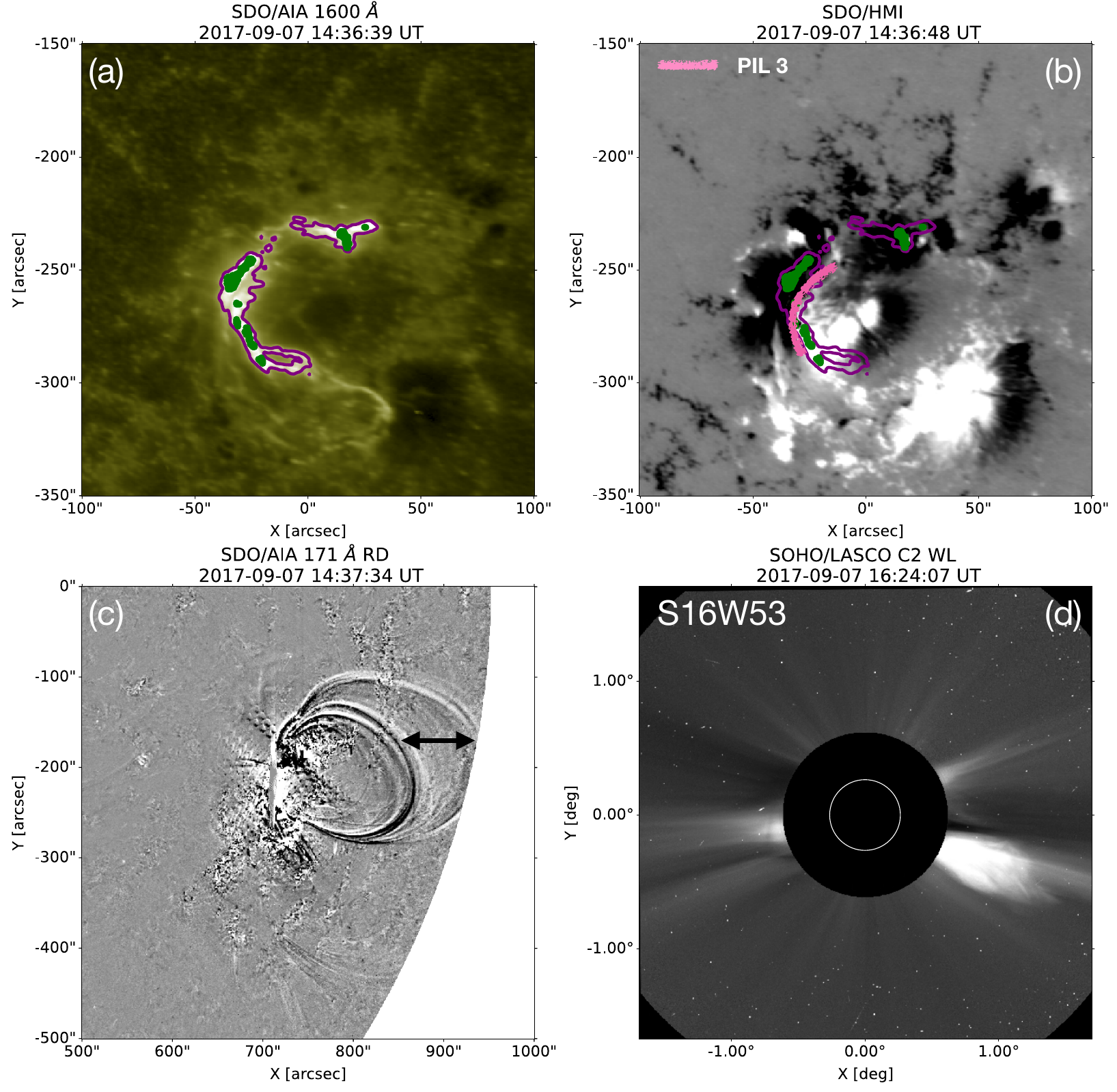}
\caption{Same as Figure~\ref{Event1} but for event four. By this time, the AR has evolved so that there is only one PIL (PIL 3, panel b). The black arrow in panel c indicates the oscillation of coronal loops rooted in the AR caused by an eruption. The online animation of panel (c) shows a movie of the SDO/AIA 171~\AA~running difference images with a 2~minute cadence between 14:01 and 15:59~UT on 2017 September 7 during the time period in which eruptive event four occurs.
}
\label{Event4}
\end{figure*}


\section{Local magnetic field configuration}

In this section, aspects of the magnetic field configuration of NOAA active region (AR) 12673, the magnetic configuration in the vicinity of the AR and the footpoints of the Earth-connected field during this time period are discussed. The decay index in the region is presented in relation to investigating CME velocity, but no detailed magnetic field analysis can be carried out since only a potential field model is used. The magnetic field configuration local to the AR (i.e. connectivity to the immediate surroundings) is analysed in order to investigate the escape of particles from the AR, which are accelerated during flare reconnection processes, and the Earth-connected fields lines are analysed in order to investigate the processes which may inject and accelerate particles towards the Earth. Two different potential field models are used for the analysis of the decay index profile and to investigate the magnetic field configuration in the vicinity of the AR. The extrapolation methods are also described in this section.

\subsection{Active region decay index height profile}

A study by \citet{kliem2021} showed that the height profile of the decay index above the critical height (i.e. the height at which the torus instability sets in for eruptive flux ropes) may correlate with CME velocity in the cases of fast CMEs (with velocity greater than 1500~km~s$^{-1}$). A similar analysis is followed here. The value of this approach is that if the correlation is found to hold, it enables a pre-event investigation of which CMEs may be of sufficiently high speed that they drive shocks as they propagate. Hence, these CMEs could be SEP-productive. Potential field models are sufficient for the analysis of the height profile of the decay index (whereas a full treatment of the AR, which contains complexity and free magnetic energy, requires the construction of a non-linear model).

Potential field models of NOAA AR 12673 are created by extrapolating linear fields using the method of \citet{alissandrakis1981} with the force-free parameter set to zero (i.e. current-free). The radial field component of photospheric magnetograms from the HMI \citep{scherrer2012} SHARP data series (Spaceweather HMI Active Region Patch; \citealt{bobra2014}) taken around the onset time of each event are used as the lower boundary of each extrapolation. The magnetograms are spatially downsized by a factor of 2, meaning that each pixel in the extrapolation volume represents approximately 0.725~Mm in each spatial dimension. The height of the extrapolation volume is chosen as 451 pixels, or approximately 327~Mm (0.47~R$_{\odot}$).

Once the potential field model has been created, the decay index is computed. The poloidal component of the magnetic field is used in the decay index calculation \citep{kliem2006, james2022}, which is approximated in this study by using the field component that is transverse to the PIL along which the flare ribbons are observed ($B_{tr} = \sqrt{(Bx^2+By^2)}$). Pixels are selected in the lower boundary of the extrapolation that correspond to the ``active" photospheric PILs associated with each flare. The mean value of the decay index along this ``active" PIL is then computed at each height layer in the extrapolation. 

The critical height, $h_{crit}$, is defined as the height of the lowest layer in the extrapolation in which the mean decay index is greater than the critical decay index, $n_{crit}$. However, any critical heights in the lowest 9 layers (3.26 Mm) of the volume are discounted as these values are generally a result of noise in the boundary magnetogram. \citet{bateman1978} found a theoretical critical decay index of 1.5, however observational and theoretical studies have determined values of $1 < n_{crit} < 2$ \citep{torok2007, fan2007, demoulin2010}. \citet{kliem2021} tested various values of the critical decay index and compared observed CME speeds to gradients of the decay index measured over different height ranges above the critical height. They found the strongest correlation between CME speeds and gradients of the decay index when the critical decay index was taken as $n_{crit}$ = 1.7 and the gradients were calculated over a range of 1-1.6 times the critical height. In this study, the values used by \citet{kliem2021} are adopted i.e., $n_{crit}$ = 1.7 and decay index gradients computed over the relative range of (1-1.6)$h_{crit}$.

The critical height $h_{crit}$ and the gradient of the decay index $dn/dh$ were calculated (see Table~\ref{tab:decay}) for the four active PILs for which flare ribbons were observed during the four eruptive events (Figures~\ref{Event1}--\ref{Event4}). At PILs 3 and 4 it can be seen that $h_{crit}$ increases and $dn/dh$ decreases with time. During events one and two, the minimum $h_{crit}$ and maximum $dn/dh$ occur at PIL 3. During event three $h_{crit}$ is lower at PIL 3 than PIL 4.

The numbers highlighted in bold in Table~\ref{tab:decay} represent $h_{crit}$ and $dn/dh$ calculated for the main PIL that was "activated" at the peak of the flare. The main PIL was chosen as the PIL above which the strongest flare ribbons were observed at the flare peak. The results show that for events one, two, three, and four (E1--E4 in Table~\ref{tab:decay}) the critical height above which a flux rope would become torus unstable, $h_{crit}$, is 39, 40, 42 and 48~Mm, respectively. Above these heights, the decay index falls with height as 0.020, 0.022, 0.017 and 0.16~Mm$^{-1}$, respectively. Following the findings of \citet{kliem2021} a correlation between CME velocity and the rate at which the decay index falls has been indicated for CMEs with speeds greater than 1500~km~s$^{-1}$, which in this study applies to the two SEP productive events only (events two and three). Our results are consistent with the findings in \citet{kliem2021} however, our values indicate that the region has the possibility of producing a fast CME ($\geqslant$1500~km~s$^{-1}$) at the time of all four events. We only observe fast CMEs during events two and three, which are SEP-productive.


\startlongtable
\begin{deluxetable*}{cccccccccc}
\tablecolumns{10}
\tablewidth{0pt}
\tablecaption{Critical height h$_{crit}$ and gradient of the critical decay index $dn/dh$ calculated for the active PILs. \label{tab:decay}}
\tablehead{
\colhead{h$_{crit}$} & \colhead{Event 1} & \colhead{Event 2} & \colhead{Event 3} &
\colhead{Event 4} & \colhead{$\frac{dn}{dh}$} & \colhead{Event 1} &
\colhead{Event 2} & \colhead{Event 3} & \colhead{Event 4} \\
\colhead{(Mm)} & & & & & \colhead{(Mm$^{-1}$)} & & & 
}
\startdata
PIL1  & 46 & 47 & - & - & & 0.022 & 0.021 & - & - \\
PIL2  & 40 & {\bf 40} & - & - & & 0.022 & {\bf 0.022} & - & -  \\
PIL3  & 25 & 25 & {\bf 42} & {\bf 48} & & 0.025 & 0.024 & {\bf 0.017} & {\bf 0.016} \\
PIL4  & {\bf 39} & 39 & 59 & - & & {\bf 0.020} & 0.020 & 0.016 & - \\
\enddata
\tablecomments{The numbers in bold represent the critical height and gradient of the critical decay index at the main active PIL at the peak of the flare during the four events (E1--4). }
\end{deluxetable*}

\subsection{Magnetic field configuration in active region vicinity}

\begin{figure*}
\centering
\includegraphics[width=1.0\textwidth]{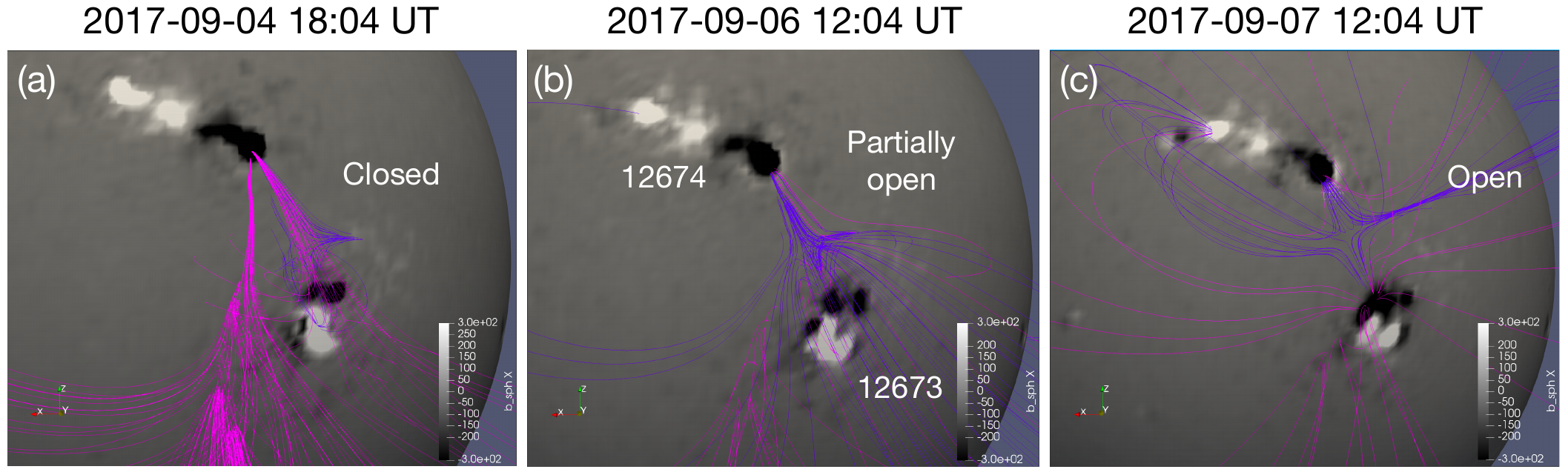}
\caption{Potential magnetic field extrapolations on September 4, 6, and 7 at 18:04~UT, 12:04~UT, and 12:04~UT, respectively. The radial magnetic field shows positive (negative) magnetic polarities that are saturated at $\pm$300~G in white (black). The pink lines represent magnetic field that is open, and the blue lines represent magnetic field that belongs to a magnetic null point located between AR 12673 and AR 12674. On September 4 (panel a) the field lines associated with the null are closed whereas, on September 6 and 7 (panels b and c) the magnetic field of the null are partially and completely open on the west side.}
\label{null}
\end{figure*}

To investigate the magnetic field configuration in the vicinity of the AR, potential field models are constructed using the PFFS model available in SSWIDL. The models are made are constructed on 2017 September 4 at 18:04~UT, 2017 September 6 at 12:04~UT, and 2017 September 7 at 12:04~UT (Figure~\ref{null}), close to the time of the four events. The models reveal the presence of a null point to the north east of the AR, between AR 12673 and AR 12674, which is located in the northern hemisphere. The null is present throughout the time period in which the four events studied here take place. The potential field models also reveal a channel of open magnetic field along the east boundary of the negative polarity, which is also present for the entire duration of the events and corresponds to a small coronal hole in the AIA observations. As can be seen in Figure~\ref{null}, the configuration of the null changes in the time period of the four events between 2017 September 4 and 7. On September 4 the null is not associated with open field lines, while on 6 and 7 the west section of the null is associated with partially and completely open field, respectively. The null is closed on the west side until September 6 due to the decayed positive magnetic flux that is close to the AR boundary and also an AR beyond the west limb.

By comparing the EUV and coronagraph data of each of the four eruptive events to the magnetic field configuration given by the potential field model the following conclusions are drawn. Event one is directed to the west of the radial and away from the null point and open magnetic field. The emission structures observed in the AIA data indicate no significant perturbation at the null. Event two, which is directed approximately radial from the AR, interacts with the null as the magnetic field expansion of the CME occurs. In event 3 the AIA EUV data shows that reconnection occurs at the null point, likely due to the CME propagation direction, which is east of the AR radial. This is also the location of open magnetic field. Finally, event four propagates to the south-west of the radial direction away from the null point and open magnetic field.

\section{Magnetic connectivity to Earth}

\begin{figure*}
\centering
\includegraphics[width=1.0\textwidth]{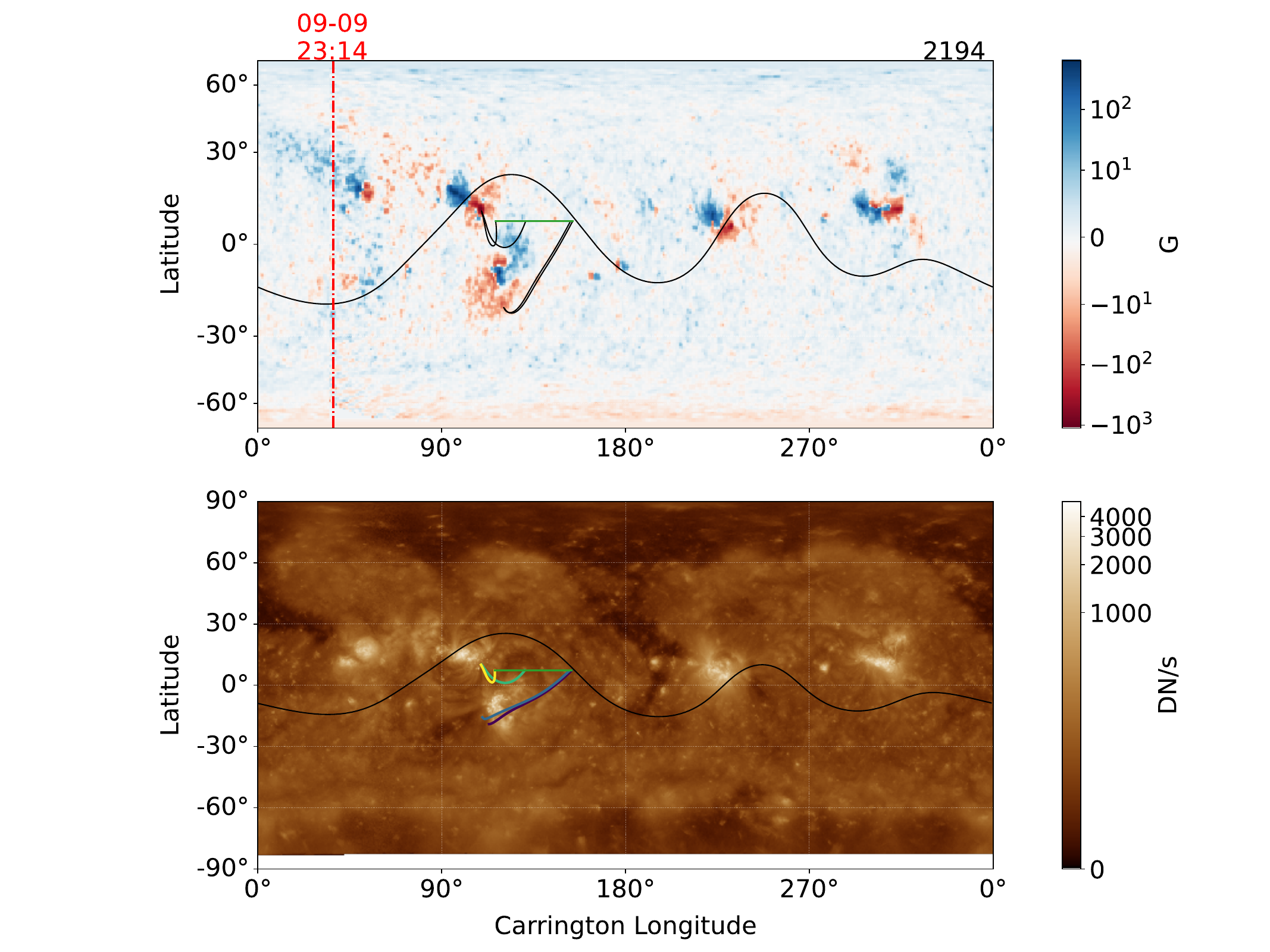}
\caption{The magnetic connectivity of Earth during 2017 September 4--7 (Carrington rotation 2194). The top panel shows the synoptic map of the longitudinal magnetic field taken by GONG on 2017 September 9 at 23:14~UT. The black lines represent the field lines connected to Earth during the time of the four eruptive events moving from right to left, and the location of the heliospheric current sheet. The green line represents the back-projected trajectory of the Earth. The grey dashed lines correspond to the Carrington times labelled above and the red dot-dashed line indicates the time and longitude that the synoptic map was last updated. In the bottom panel the corresponding SDO/AIA 193~\AA~synoptic map is shown. The black and green lines correspond to position of the heliospheric current sheet and back-tracked trajectory of as above. The purple, blue, green, and yellow lines show the magnetic field lines connected to Earth on 2017 September 4 at 18:00~UT, 2017 September 4 at 20:05~UT, 2017 September 6 at 11:52~UT, and 2017 September 7 at 14:31~UT, respectively.
}
\label{earth_connectivity}
\end{figure*}

For energetic particles to escape and reach Earth they ultimately need to be injected onto open magnetic field lines that are magnetically connected to Earth. To investigate the footpoints of open field that is magnetically connected to Earth during 2017 September 4--7 we use a combination of the potential field source surface model (PFSS, \citealt{schatten1969}) along with a ballistic propagation model \citep{Neugebauer1998}.

We use a Global Oscillations Network Group (GONG) synoptic photospheric magnetic field map taken on 2017 September 9 at 23:14~UT to construct the PFSS model. The GONG map is loaded into Python using SunPy \citep{sunpy2020} and pfsspy \citep{yeates2018, stansby2019} was used to construct the potential field between 1~R${_\odot}$ and 2.5~R$_{\odot}$. Then a Parker spiral configuration is assumed above 2.5~R$_{\odot}$ using a solar wind speed of 500~km~s$^{-1}$. This speed was the average solar wind speed measured by ACE during our time period. We then use HelioPy \citep{stansby2021}, along with SpiceyPy \citep{annex2021} and the SPICE toolkit \citep{acton2018} to trace the field lines connected to Earth back to their source on the surface. We do this for four different times (2017 September 4 at 18:00~UT, 2017 September 4 at 20:00~UT, 2017 September 6 at 11:52~UT, and 2017 September 7 at 14:31~UT) at the start of each eruptive event (see Section~\ref{sec:events}). We choose these times so that we can determine the instantaneous Earth connectivity and estimate the source location of the energetic particles measured near-Earth.

Figure~\ref{earth_connectivity} shows the magnetic connectivity of Earth at the times of the four eruptive events. The top panel shows the Earth's back-projected trajectory (green line), the traced field lines (black lines), and the location of the heliospheric current sheet (black line) overlaid on the GONG synoptic magnetogram. The bottom panel shows the same but overlaid on a SDO/AIA 193~\AA~synoptic map, constructed by joining 27 AIA images together, with the final image taken on 2017 September 10.

The results show that for the first two events (on 2017 September 4) the Earth-connected field lines are rooted to the south-east of AR 12673 in decayed negative polarity field. We recall that the first event (2017 September 4 at 18:05~UT) was non-SEP productive, and the CME propagated to the south-west. The second event (2017 September 4 at 20:00~UT), which was SEP productive, propagated radially from AR 12673. However, it was a wide CME, likely to expand into Earth-connected field lines to the south-east of the source AR.

For events three and four, the connectivity has changed, and Earth-connected field lines are rooted in NOAA AR 12674 (to the north-east of 12673). In particular, the field lines are rooted in the main positive polarity spot of AR 12674. The magnetic field from this part of the AR forms part of the null point that exists between ARs 12673 and 12674. As we have seen in Section~\ref{sec:events}, the erupting loop structures in event three propagate in many directions, including into the null, and where AR 12674 is connected to Earth. In event four, the eruption propagates to the west away from the region that is well-connected to Earth.


\section{Summary \& Discussion}

\begin{figure*}
\centering
\includegraphics[width=1.0\textwidth]{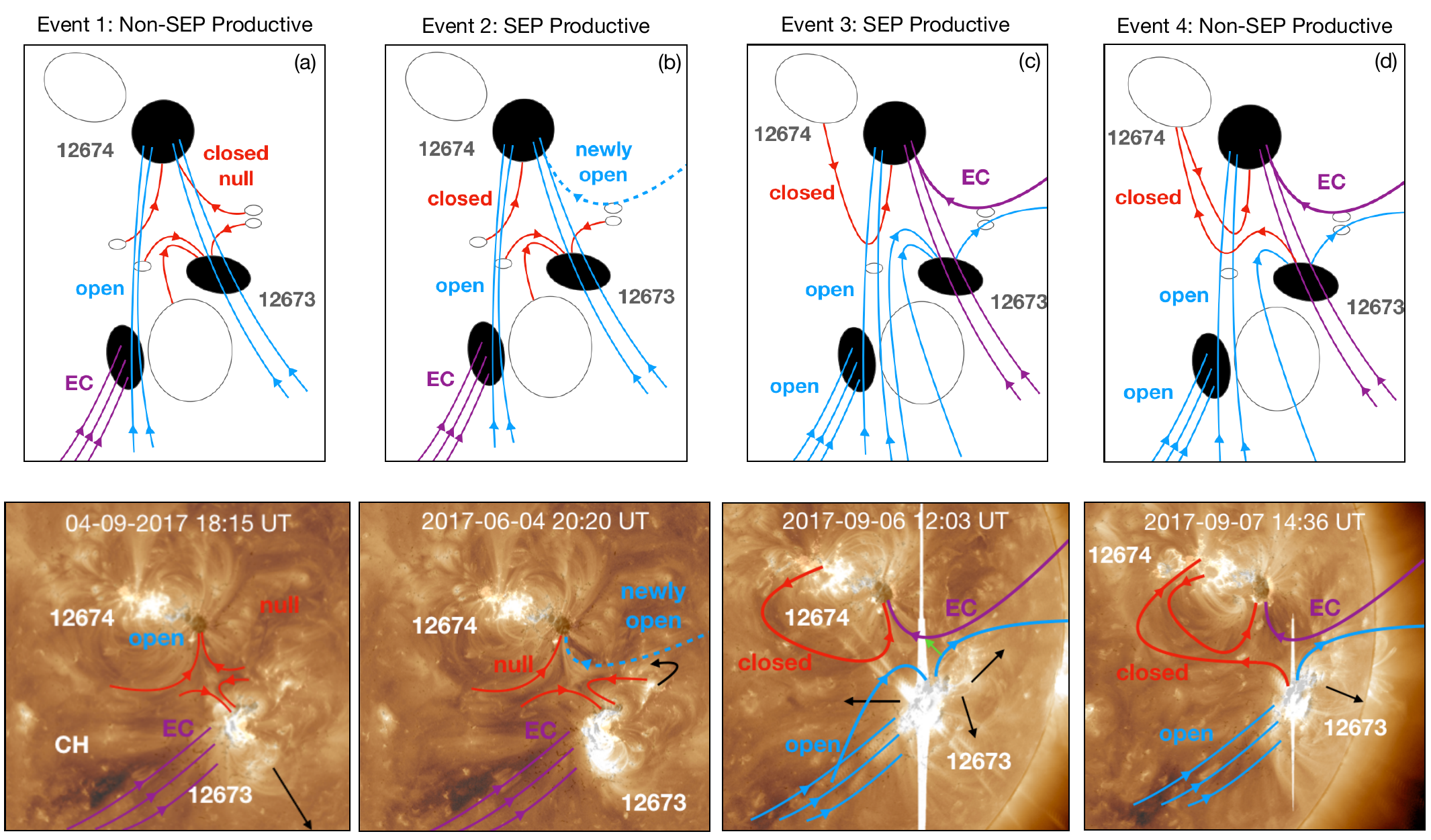}
\caption{A schematic showing the magnetic field configuration (null point, open magnetic field, Earth-connected field), taken from the PFSS model, at the time of each eruptive event along with the 193~\AA~emission and photospheric magnetic field. In the top panels the white (black) features represent the positive and negative magnetic polarities of ARs 12673 and 12674 along with decayed/quiet Sun field. Red lines show the location of closed magnetic field whereas, blue represents open magnetic field. Purple shows the magnetic field lines that are Earth-connected (EC), taken from the results of the PFSS and ballistic propagation model. In the bottom panels the same labels apply. There is a small coronal hole (CH) visible to the east of AR 12673. The black arrows show the propagation directions of the eruptions. Finally, in panel (c) the green arrow demonstrates the direction of the coronal loops during event three that drives reconnection at the null.}
\label{cartoon}
\end{figure*}

The analysis presented here uses AR 12673 as a case study to investigate whether and how the magnetic environment of an AR plays a role in flare/CME events being (or not being) SEP productive. Several aspects of the magnetic field are considered including how the strength of the field falls with height in the AR, and whether that correlates with CME speed (important for generating shocks), how the magnetic field of the AR interacts with its surrounding field during the dynamic phase of the flare/CME, and how particles may get injected onto observer-connected field lines. Despite the four events occurring close in time, there are some interesting differences and important findings that are summarised here and in Figure~\ref{cartoon}.

In event one, which was non-SEP productive, a major flare occurred (M1.0 GOES-class), and a relatively fast CME that had a radial speed of 973 km~s$^{-1}$. At this time, the Earth-connected field lines were rooted in a region of negative polarity field to the south-east of NOAA AR 12673 (see purple lines in Figure~\ref{cartoon}~a), not in the core of the AR, where the flare ribbons indicate magnetic connection to locations of particle acceleration due to flare processes. The CME occurring during this event was fairly narrow and propagated to the west of AR 12673's radial direction, i.e. away from earth-connected field lines (black arrow in Figure~\ref{cartoon}~a). Perhaps deflected by the small coronal hole (CH) to the east of the AR. No perturbation of the null was observed in EUV imaging data, in line with the CME propagation direction (i.e. away from the null), and no CME-driven shock was produced. The data indicate that flare accelerated particles were not able be re-directed via magnetic reconnection onto Earth-connected field lines, and no shock acceleration of particles occurred due to the CME propagation.

Event two was SEP productive and involved a flare and CME that was initiated less than two hours after event one. Very little evolution of the AR corona took place during this short interval, in terms of photospheric motions and flux emergence, and the location of the Earth-connected field lines remained the same at the Sun as it was for event one (purple lines Figure~\ref{cartoon}~b). What was significantly different were the characteristics of the CME, which expanded to have an angular width of $\sim$200$^{\circ}$, and propagated into an environment already modified by the previous CME (event one). This expansion appears to have been sufficient to cause the CME to interact with the Earth connected-field lines. The shock created by this CME likely accelerated particles along the open field lines, meaning the particles were able to reach Earth. The expansion of the CME also activated the null between ARs 12673 and 12674 (red and dashed blue lines in Figure~\ref{cartoon}~b), leading to reconnection. However, this reconnection (and any transfer of particles) did not involve any open field lines that were Earth-connected. Collectively, these observations are suggestive of particle acceleration occurring along Earth-connected open field to the south-east of AR 12673, accelerated by the CME shock, consistent with a post-eruption origin as found by \citet{chertok2018a,chertok2018b}.

At the time of event three, the location of the Earth-connnected field lines had moved from its previous position to the south-east of AR 12673 and into the negative polarity (leading) spot of AR 12674 (purple lines Figure~\ref{cartoon}~c). The CME that occurred as part of event three, propagated to the north-east, radially from the AR (black arrows in Figure~\ref{cartoon}~c), and therefore towards the null. An activation of the null was evidenced in EUV imaging data and reconnection at the null effectively opened field lines in AR 12763 (as it transferred the footpoints of the open Earth-connected field from AR 12674 to AR 12673), providing a magnetic channel for flare accelerated particles to escape to Earth. In addition, the detected type II burst indicates the occurrence of a CME-driven shock that also may have accelerated particles along Earth-connected field lines. It is interesting to note that the SEP energy spectrum, although still soft, contained higher energy protons (of a few hundred MeV, see Figure~\ref{hmi_goes} b) than were detected at Earth during event two \citep{bruno2019}.  

The CME of event four was narrow and was deflected to the south-west, away from Earth connected field lines that remained in NOAA AR 12674 at this time, albeit modified by the null-point reconnection of event three (see Figure~\ref{cartoon}~d). Although the null point was well developed at this time, there appears to have been no activation of the null (i.e. reconnection). The relatively small and slow CME, deflected away from Earth-connected field lines, with no reconnection to transfer particles, seems to be at the heart of why event four was not SEP-productive.

The previous work of \cite{chertok2018b} analysed the CMEs of the two SEP-productive events (i.e. our events two and three) including the propagation of the CMEs through the high-speed solar wind stream emanating from the coronal hole to the south-east of AR 12673. Their Figure 1 shows that dimmings associated with the field expansion of the CMEs in these two events extends to AR 12674. This supports our observational findings that reconnection at the null (activation of the null) occurs in both SEP-productive events, transferring the footpoints of some of the erupting magnetic field from AR 12673 to AR 12674. However, this reconnection, and the new magnetic pathways it creates, likely only becomes significant for the second SEP-productive event (our event three), since only from this time on are Earth-connected field lines rooted in AR 12764. It could be speculated that the scenarios of CME-shock accelerated particles along Earth-connected field (event two) and flare-processes, as well as shock accelerated particles (in event three), could contribute to the different energy spectra observed in situ. In that respect, protons arriving at Earth from event two are limited to energies below 150 MeV whereas for event three the protons reach energies of a few hundred MeV \citep{bruno2019}. Indeed, \cite{bruno2019} find that the temporal evolution of the SEP events are complex but conclude that both events show evidence of CME-shock accelerated particles.

When modelling the magnetic field of an AR, non-linear force free field models are usually more appropriate than potential field models as ARs consist of non-potential field configurations with varying degrees of shear and twist across the configuration. However, in this study we are interested in the large-scale magnetic field of the solar corona surrounding the AR where the magnetic field is known to be close-to a potential state and the decay index profile of the magnetic field overlying the AR, which only requires knowledge of the potential field. To validate the PFSS model for the field surrounding the AR, we used EUV emission structures observed by SDO/AIA. For example, the field associated with the magnetic null, which is observed to the north-east of AR 12673, and the small coronal hole to the east where open magnetic field is located. 

Very strong magnetic fields have been recorded in this AR \citep{wang2018} and therefore numerous artifacts in the line-of-sight and vector magnetic field exist \citep{anfinogentov2019}. These artefacts are present along one of the AR’s PILs and are most apparent in the tranverse field component. Therefore, our approach of using the PFSS model, involving the radial magnetic field component as the boundary condition, our investigation of the magnetic field in the vicinity of the AR, and the gradient of the overlying magnetic field with height involve field lines with footpoints away from the PILs in the AR. Therefore, these artefacts do not affect our analysis of the configuration of the magnetic field provided by the potential field extrapolations.

Using a potential field model allows the rate at which the downward Lorentz force of the overlying field varies with height to be determined. \cite{kliem2021} have shown that the rate at which this force varies with height is correlated with CME speed for CMEs with speeds greater than 1500~km~s$^{-1}$. Both SEP-productive events in this study have CMEs with speeds greater than this value, whereas the non SEP-productive events are slower. Using the same height range as \cite{kliem2021} we calculated the change of the decay index with altitude $dn/dh$, above the critical height $h_{crit}$ for all four events. We obtained the change of the decay index with altitude above the flare/CME sites (as indicated by the location of the flare ribbons) in order to determine whether this metric indicates when an AR might be capable of producing a fast CME and therefore creating shock-accelerated particles. Our results show little variation in $dn/dh$ at the times of the four events, even though two of the CMEs have speeds below 1500~km~s$^{-1}$. Event three shows a lower $dn/dh$ than expected as the region produces a CME that is $>$2000~km~s$^{-1}$. However, event three is our most complex event and the flare seemingly has two parts; confined and eruptive. Flare ribbons are seen to extend along PIL 3 and then 4 by the peak of the flare. The critical heights are quite different (42 vs 59~Mm above PILs 3 and 4), for this event however, the values of $dn/dh$ are very similar (0.017 and 0.016~Mm$^{-1}$). The complexity of the events produced by this region brings into question, which PILs should be used for the calculation of $h_{crit}$ and $dn/dh$. Nevertheless, this is a very preliminary study in a complex AR with several PILs activating during each event and more analysis is required in order to determine whether the parameters $h_{crit}$ and $dn/dh$ could be interesting proxies to consider alongside other characteristics of an AR's magnetic field when assessing likelihood of SEP occurrence.


\begin{acknowledgments}
SLY and LMG would like to thank NERC for funding via the SWIMMR Aviation Risk Modelling (SWARM) project (grant no. NE/V002899/1). SLY would also like to thank DIPC for their hospitality at Miramar Palace. AWJ is supported by a European Space Agency (ESA) Research Fellowship. DS received support under STFC grant number ST/S000240/1. TM is supported by the STFC PhD studentship grant ST/V507155/1.
\end{acknowledgments}

\vspace{5mm}
\facilities{SDO/AIA, SDO/HMI, GONG, SoHO/LASCO, GOES}

\software{JHelioviewer \citep{muller2017},  
          SunPy \citep{sunpy2020}, 
          pfsspy \citep{yeates2018, stansby2019},
          HelioPy \citep{stansby2021},
          SpiceyPy \citep{annex2021},
          SPICE toolkit \citep{acton2018},
          }

\bibliography{ref}{}
\bibliographystyle{aasjournal}

\end{document}